\documentclass[aps, llncs, twocolumn]{revtex4-1} 
\usepackage{graphicx}
\usepackage{amsmath}
\usepackage{float}
\begin{document}
\title{Odd-even rule for zero-bias tunneling conductance in coupled Majorana wire arrays } 

\author{Deepti Rana}
\author{Goutam Sheet}
\email{goutam@iisermohali.ac.in}
\affiliation{Department of Physical Sciences, 
Indian Institute of Science Education and Research
Mohali, Mohali, Punjab, India}

\begin{abstract}

A semiconducting nanowire with strong Rashba coupling and in proximity of a superconductor hosts Majorana edge modes. An array of such nanowires with inter-wire coupling gives an approximate description of a two-dimensional topological superconductor, where depending on the strength of the magnetic field and the chemical potential, a rich phase diagram hosting trivial and different types of non-trivial phases can be achieved. Here, we theoretically consider such a two-dimensional assembly of spin-orbit coupled superconducting nanowires and calculate the collective tunneling conductance between normal electrodes and the wires in the topological regime. When the number of wires in the assembly is $N$, as a consequence of the way the Majorana bonding and anti-bonding states form, we find that $N$ conductance peaks symmetric about the bias $V = 0$ appear, for even $N$. When $N$ is odd, a ZBCP also appears. Such an assembly can be realized by standard nano-fabrication techniques where individual nanowires can be turned $ON$ or $OFF$ by using mechanical switch (or local top gating) to make $N$ either even or odd -- thereby switching the ZBCP $OFF$ or $ON$, respectively. Hence, our results can be used to realize and detect topological superconductivity efficiently, unambiguously and in a controlled manner.

\end{abstract}
\maketitle

A topological superconductor is a superconductor for which a non-zero superconducting (pairing) gap exists in the bulk while the boundary hosts gapless self-hermitian modes \cite{Hasan, Qi, Ando, Alicea(a), Beenakker}. Because of their self-hermitian properties, such boundary modes are termed as the "Majorana modes". In a one-dimensional (1D) topological superconducting system, like a superconducting, spin-orbit coupled quantum wire (a Majorana wire) \cite{Lutchyn, Oreg, Mourik, Deng, Rokhinson, Anindya, Finck, Churchill, Chang}, these zero-energy modes are bound to the ends of the wire \cite{Kitaev(a)}. In this context these modes are also often referred to as Majorana zero modes or Majorana bound states. In a two dimensional (2D) topological superconductor, the boundary hosts gap-less Majorana chiral modes which propagate along the edge \cite{Read}. On the other hand, the cores of quantum vortices in the bulk of such 2D superconductors host Majorana bound states \cite{Alicea(b), Fu, Nagaosa}. Like all superconductors, such topological superconductors also respect robust particle-hole symmetry -- as a consequence of this, the boundary modes are also extremely robust. The physics of Majorana modes have attracted substantial attention of the contemporary condensed matter physics community for their exotic fundamental properties like, their non-abelian exchange statistics, which makes them potentially important ingredients for building a topological quantum computer \cite{Kitaev(b), Sarma(a), Nayak, Leijnse,  Stanescu, Elliot, Sarma(b)}. 

In this paper, we theoretically considered an assembly of Majorana nanowires placed on an $s$-wave superconductor \cite{Seroussi}. We calculated the tunneling conductance between normal electrodes mounted on the wires and the assembly (array) of the wires. Our key observations are: (a) when the number of wires in the assembly is $N$, and $N$ is even, $N$ conductance peaks symmetric about the bias $V = 0$ appear, (b) when $N$ is odd, a ZBCP, along with the conductance peaks symmetric about $V = 0$ also appears. These are subject to certain condition that will be discussed in detail later. It is known that mere observation of a ZBCP does not provide a solid signature of Majorana bound states. This is mainly because a ZBCP in tunneling spectroscopy involving superconductors can appear for a number of reasons other than Majorana bound states \cite{Liu, Cho, Pikulin}. As per our calculations, simply by changing the number of transport-active wires (e.g., through mechanical switch or by local top gating) it will be possible to probe the Majorana states unambiguously through controlled appearance and disappearance of the ZBCP by making $N$ odd and even, respectively.

Our model setup is shown schematically in Figure 1 (a). The setup consists of an array (in the $x-y$ plane) of $N$ parallel Rasbha nanowires, lying in close proximity to an $s$-wave superconductor, with Zeeman field ($V_z$)  applied along the $\hat{z}$ direction. As a consequence of superconducting proximity effect, a superconducting gap ($\Delta$) is induced in the wires. The model also includes inter-wire tunneling which is facilitated by the underlying superconductor. 
 
  The  Hamiltonian for the system described above can be written as :
 \[ H = H^{||} + H^\perp\tag{1} \]
where $H^{||}$  is the Hamiltonian describing intra-wire dynamics while and $H^\perp$ describes  the inter-wire dynamics. In the mean field approximation, these two Hamiltonians can be expressed as:
\[H^{||}=-t_x\sum_{i,\delta, \sigma}[\ c^\dagger  _{i+\delta_x\sigma} c_{i\sigma}] -\mu\sum_{i\sigma} c^\dagger_{i\sigma}c_{i\sigma}-V_z\sum_i[ c^\dagger_{i\uparrow}c_{i\downarrow}+\]\[h.c.] +\Delta\sum_i[c^\dagger_{i\uparrow}c^\dagger_{i\downarrow}+c_{i\downarrow}c_{i\uparrow}]+\frac{i\alpha}{2}\sum_{i,\delta}[ c^\dagger_{i+\delta_x}\sigma_y c_i+h.c.]
\tag{2}  \]

\[H^\perp = -t_\perp\sum_{i,\delta,\sigma} [c^\dagger_{i+\delta_y\sigma}c_{i\sigma}+h.c.]-\frac{i\beta}{2}\sum_{i,\delta}[c^\dagger_{i+\delta_y}\sigma_x c_i+h.c.]\tag{3}
\]
where $c^\dagger_{i\sigma}(c_{i\sigma})$ is the fermionic operator that creates (annihilates) a particle at lattice site $i=\{i_x,i_y\}$ with spin $\sigma$. $\delta_x$ and $\delta_y$ represent the nearest neighbour lattice vectors in $x$ and $y$ directions respectively. $t_x$ gives the hopping matrix element along the wires (chosen to be along $x$),  $\alpha$ is the Rasbha spin-orbit coupling , $\mu$ is the chemical potential, $t_\perp$ is the weak inter-wire hopping matrix element, $\beta$ is the weak spin-orbit coupling term between the wires in transverse direction and that is linked to inter-wire hopping. It is understood that the effective inter-wire coupling between the wires can be significant only when the distance between the wires is less than a coherence length in the superconductor underneath.

The Hamiltonian (1) can be written in the momentum space as :\[H=\frac{1}{2}\sum \Psi^\dagger_k h(k) \Psi_k \ ;\ k=(k_x,k_y)\] \[h(k)=\epsilon_k\tau_z+\alpha sin(k_x)\tau_z\sigma_y- V_z \sigma_z+\Delta\tau_x+\beta sin(k_y) \tau_z \sigma_x\tag{4}\]
where $\epsilon_k=-2t_xcos(k_x)-2t_\perp cos(k_y)-\mu$, $\sigma$ and $\tau$ are Pauli matrices in spin space and particle-hole basis respectively. The uncoupled wires ( $t_\perp$=0; $\beta$=0) can be tuned into two phases: (a) trival phase, when $V_z<\sqrt{\Delta^2+(\mu+2t_x)^2}$ and (b) topological phase, when $V_z>\sqrt{\Delta^2+(\mu+2t_x)^2}$. Here, $\mu+2t_x$ is the chemical potential measured w.r.t bottom of the conduction band. We have chosen $t_x=t$ as the energy unit for our calculations. 

Now, the Hamiltonian can be written as:

\[H=\sum_j\int dk_x[\epsilon_j(k_x)\Psi^\dagger_{k_x,j}\Psi_{k_x,j}+\alpha sin(k_x)\Psi^\dagger
_{k_x,j} \sigma_y\Psi_{k_x,j}\]
\[ -V_z\Psi^\dagger_{kx,j}\sigma_z\Psi_{k_x,j} + \Delta \Psi^\dagger_{k_x,j}(i\sigma_y)\Psi^\dagger_{-k_x,j}+h.c.]
\]
\[+\sum_j\int dk_x[-t_\perp\Psi^\dagger_{k_x,j}\Psi_{k_x,j+1}-i\beta\Psi^\dagger_{k_x,j}\sigma_x\Psi_{k_x,j+1}+h.c.]\tag{5}\]
where $\epsilon_j(k_x)=-2t_xcos(k_x)-\mu$
\begin{figure}[h!]
 \centering
  \includegraphics[scale=0.2]{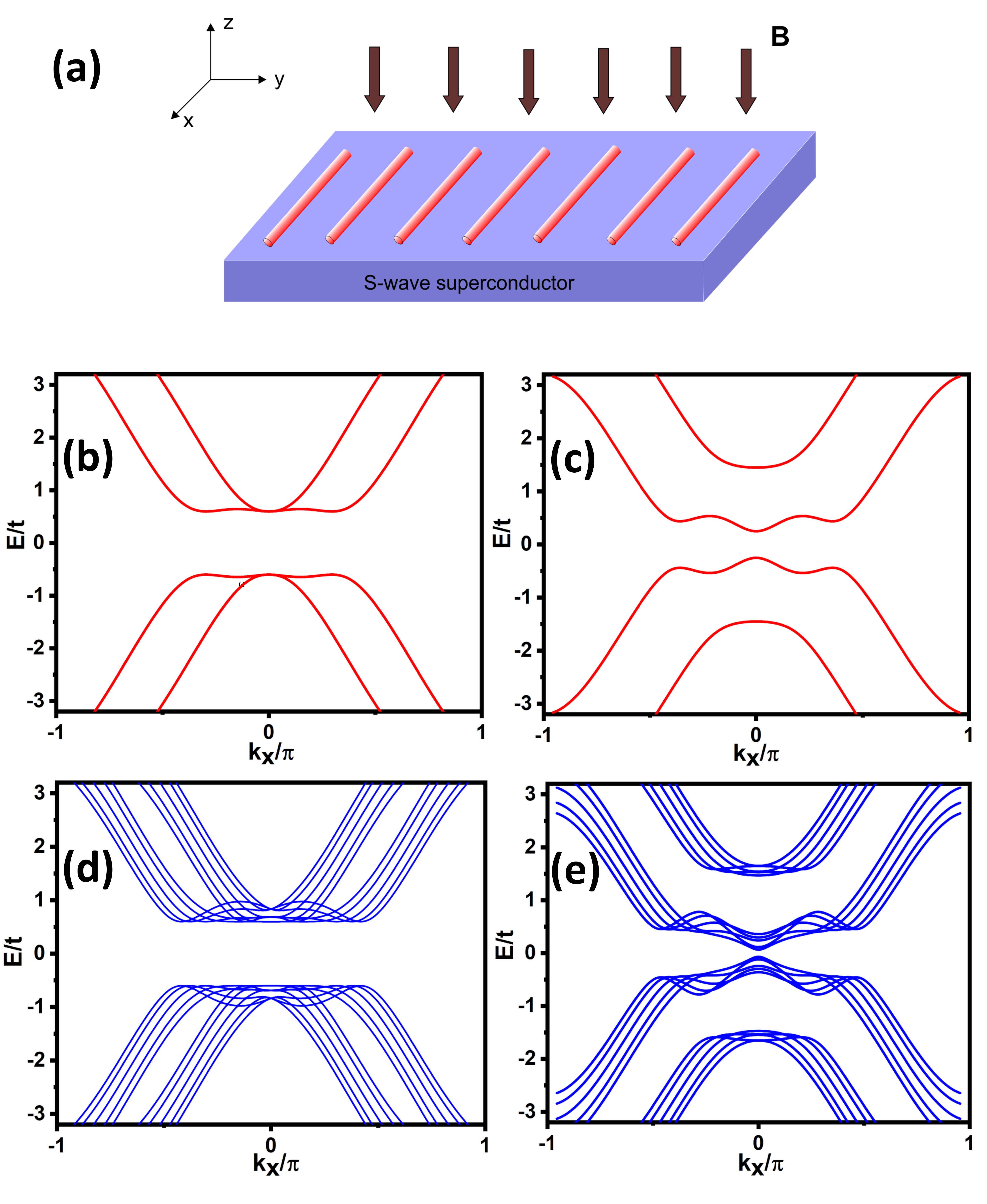}

	\caption{{(a) Schematic illustrating an array of parallel Rasbha nanowires lying in proximity to s-wave superconductor and with Zeeman field $V_z$ applied along $z$ direction. Energy dispersion for an assembly of  five Majorana wires as a function of the momentum along the wires, ($k_x$): for uncoupled (in red) (b) trivial regime ($V_z$ = 0, $t_\perp$ = 0, $\beta$ = 0) (c)  topological regime ($V_z$ = 0.85$t$, $t_\perp$ = 0, $\beta$ = 0)
	and  coupled (in blue) (d) trivial regime ($V_z$ = 0, $t_\perp$ = 0.3$t$, $\beta$ = 0.3$t$)  (e) topological regime ($V_z$ = 0.85$t$, $t_\perp$ = 0.3$t$, $\beta$ = 0.3$t$) is shown. The other parameters used for calculations are $\alpha$ = 1$t$, $\mu$ = -2$t$ and $\Delta$ = 0.6$t$ 
  }}
\end{figure}

From this, the spectrum of the system is obtained as :
\[E^2=V^2_z+\Delta^2+\epsilon^2_k+|\gamma_k|^2\pm2 \sqrt{(V_z\Delta)^2+(V_z^2+|\gamma_k|^2)\epsilon_k^2}\tag{6}\]
where $\gamma_k=\alpha sin(k_x) i+\beta sin(k_y).$

 The energy dispersion  as a function of $k_x$ is calculated for the  trivial case when $V_z=0$ for an assembly of five  uncoupled (Figure 1 (b)) and weakly coupled (Figure 1 (d)) wires. For $V_z$= 0.85$t$ the system is in topological regime. In this regime, the dispersion of  five uncoupled (Figure 1 (c)) and weakly coupled wires (Figure 1(e)) are also shown. Periodic boundary conditions along $x$ direction and open boundary conditions along $y$ direction are enforced that makes the momentum along $x$ a good quantum number. The results presented in Figure 1 are consistent with the earlier calculations on coupled Majorana wires \cite{Qu}.

In Figure 2, we show the variation of the energy spectra with the Zeeman field $V_z$. Figure 2 (a) and 2 (b) show the variation for an assembly of 3 and 4 uncoupled wires respectively while Figure 2 (c) and 2 (d) show the variation for an assembly of 3 and 4 coupled wires respectively, in the topological regime. When the wires don't have any inter-wire coupling ($t_\perp=0; \beta=0$) , as expected, it is seen that for $V_z> \Delta$ ($\Delta$= 0.6$t$ in this case), we obtain zero energy states indicating the topological regime. Now, when there is non-zero interwire coupling ($t_\perp \neq 0;  \beta\neq 0$), the spectra get modified due to mixing of states and the topological regime is achieved for a higher value of Zeeman field ($V_z > 0.75t$ in this case). Futhermore, it is also observed that for three wires, states at $E=0$ appear along with states that emerge at finite energy near zero energy symmetric about $E=0$. On the other hand, for a 4-wire assembly no zero energy states can be seen but 4 peaks around $E=0$ are seen. We will later see that these lead to odd-even rule in the conductance which is the focal theme of this article.

\begin{figure}[h!]
 \centering
  \includegraphics[scale=0.29]{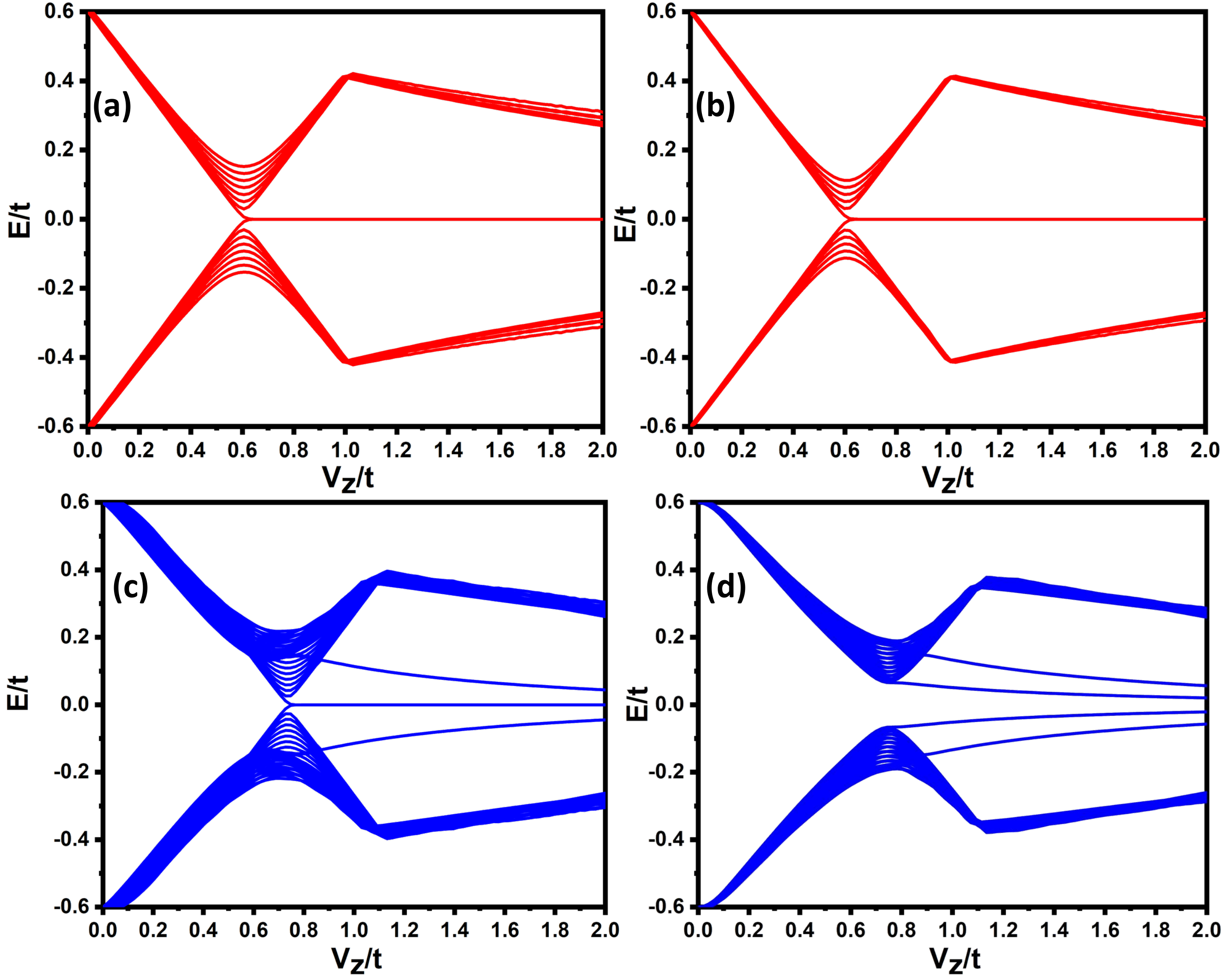}

	\caption{{Plot of energy spectrum as a function of Zeeman field in the topological regime for : (a), (b)  $N$= 3  and 4 uncoupled (in red) wires respectively, both hosting zero energy states beyond $V_z$= 0.6t (c), (d) $N$= 3 and 4 coupled (in blue) wires where, for $N$= 3, a zero energy state can be seen beyond $V_z > 0.75t$ while no such state appears for $N$=4. Number of states near zero energy is equal to $N$ in both the cases. Lowest 48 levels have been plotted. The parameters used for plotting for both uncoupled and coupled case in the topological regime are mentioned in Figure 1.
  }}
\end{figure}

\begin{figure}[h!]
 \centering
  \includegraphics[scale=0.17]{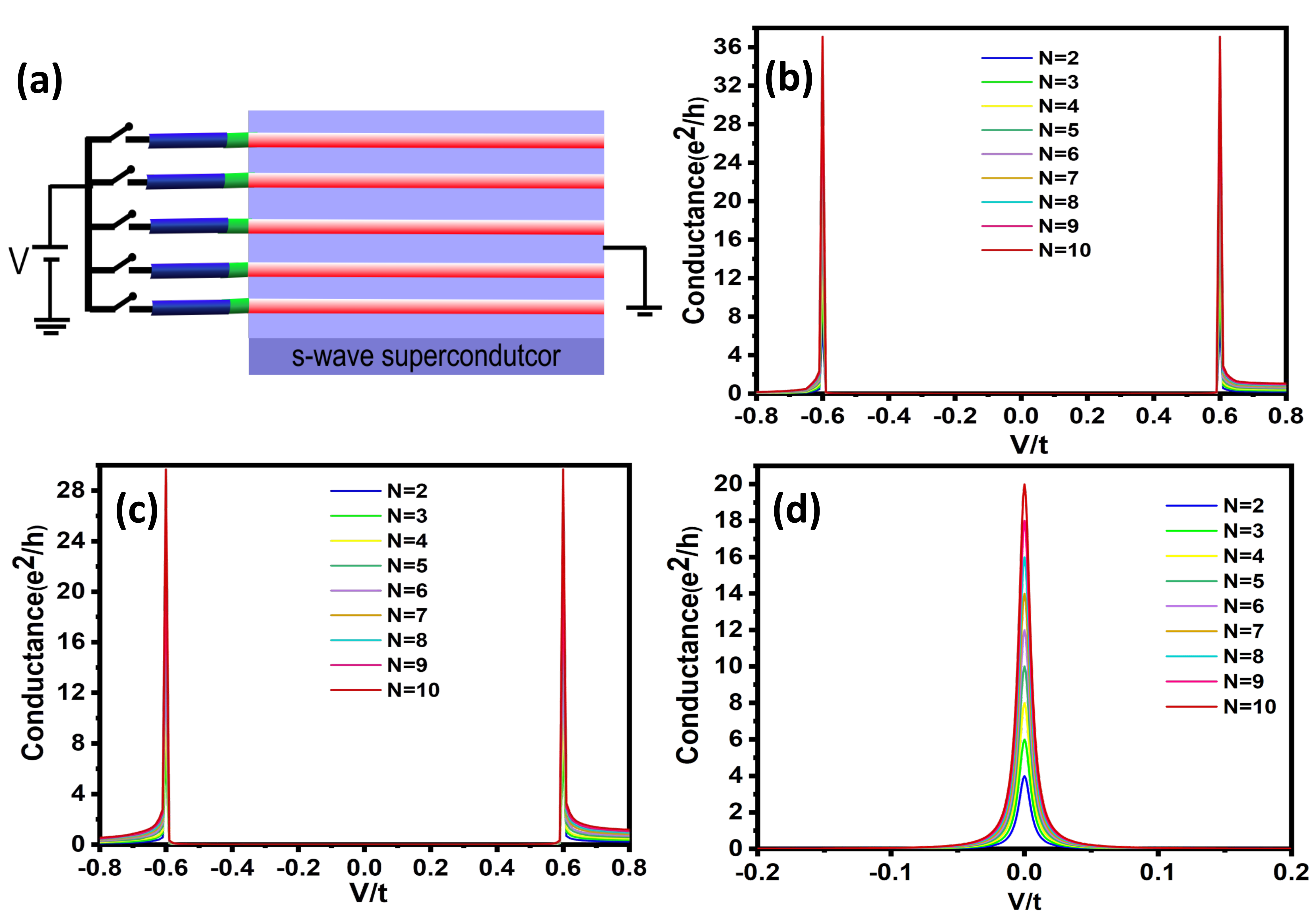}

	\caption{{(a) Schematic for measuring tunneling conductance of array of Majorana wires. The normal leads (in blue) are connected to Majorana wires (red) with a tunnel barrier (green) at the interface of the junction. A bias $V$ is applied to the normal leads. Using mechanical switch, the number of transport active wires can be controlled. Plot of zero temperature differential conductance vs applied bias $V$ of different number of  wires for (b) trivial uncoupled case (c) trivial coupled case (d) topological uncoupled case.}}.
\end{figure}

Now, we design a thought tunneling experiment to investigate the possible role of inter-wire coupling in transport through the aforementioned Majorana wire assembly. A schematic  representation of the said set-up is shown in Figure 3 (a). A normal lead is attached to one end of each semi-infinite semiconducting nanowire with strong Rasbha coupling proximitized by s-wave superconductor in the presence of an external applied magnetic field. The interface between the metal electrode and each wire falls in the tunneling regime of transport. The normal lead has the same Hamiltonian (equation 2) as the nanowire except for the superconducting term ($\Delta$). Also, chemical potential of the normal lead is chosen greater than superconducting gap so that the normal lead remains topologically trivial. The tunnel barrier at the interface of each N-S junction is modelled by adding an additional onsite energy of strength 10$t$ on one end (left) of each wire. We have assumed semi-infinite nanowires for our calculations to exclude finite size effects \cite{Sarma(c)}. 
\begin{figure}[h!]
 \centering
  \includegraphics[scale=0.3]{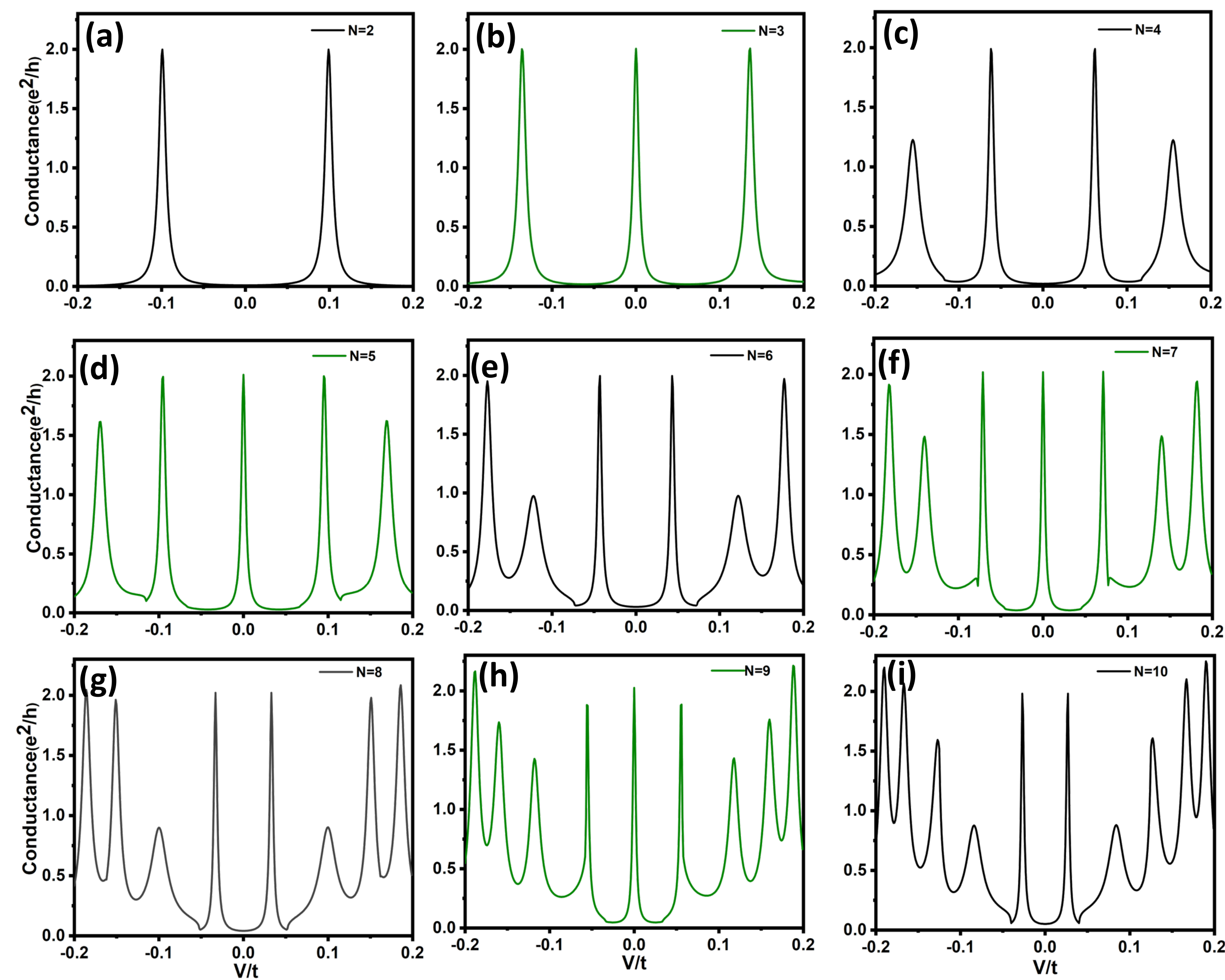}

  \caption{{Plot of differential conductance vs applied bias for $N$ = 2-10 Majorana wires for coupled topological regime ($V_z$= 0.85$t$; $t_y$ = 0.3$t$; $\beta$= 0.3$t$). The ZBCP peak appears only in the case of conductance of odd number of wires (shown in green) while no such peak is observed in the case of conductance of even number of wires (shown in black).
  }}
	
\end{figure}

 The zero temperature tunneling conductance of the array is calculated using $S$-matrix method. By computing the reflection matrix at the N-S junction, tunneling conductance can be found; where the reflection matrix ($r$) is expressed in terms of electron and hole scattering channels at energy $E$ as :
\[r=\begin{pmatrix}
r_{ee}& r_{eh}\\
r_{he}& r_{hh} 
\end{pmatrix}\tag{7}\]
where $r_{ee}$ $(r_{eh}) $ refers to the normal (Andreev) reflection submatrix. For $N$ conducting channels in the lead the  differential conductance can be evaluated using the Blonder-Tinkham-Klapwijk (BTK) formula \cite{Blonder} (in the units of $e^2/h$):
\[G=dI/dV=[N-Tr(r_{ee}r^\dagger_{ee}-r_{eh}r^\dagger_{eh})]_{E=V} \tag{8}\]
We have employed KWANT, a numerical transport package in Python, to calculate the components of the reflection matrix \cite{Groth}.

 Figure 3 (b) shows the differential conductance vs applied bias $V$, when there is no inter-wire coupling in the non-topological (trivial) regime . The differential conductance resembles the zero temperature  density of states of a conventional superconductor with coherence peaks appearing at induced superconducting gap energy $\pm\Delta$. As expected, the magnitude of conductance at $\pm$ $\Delta$ increases monotonically with increase in number of wires. Figure 3 (c) depicts the  differential conductance when there is inter-wire coupling in the non-topological (trivial) case for $N$ number of wires. Unlike the trivial uncoupled case, enhancement of magnitude of conductance at $\pm$ $\Delta$ is not monotonic with increase in number of wires. Figure 3 (d) shows the plot of conductance vs applied bias $ V$ for uncoupled topological phase where each Majorana bound state contributes a conductance of $2e^2/h$ to the total differential conductance. Upto this part, the number of wires only contributes to the overall scaling of the absolute magnitude of the differential conductance. However, more interesting physics emerges when weak coupling between the wires is also introduced in the topological regime.

As it can be seen in Figure 4, in the topological regime, turning on weak coupling between the wires has resulted in  $N$ conductance peaks, symmetric about bias $V$=0. The most important feature here is the emergence of a ZBCP when $N$ is odd (Figure 4 (b), 4 (d), 4 (f) and 4 (h)). On the other hand, for even values of $N$ (Figure 4 (a), 4 (c), 4 (e), 4 (g) and 4 (i)), ZBCP doesn't appear.  This is due to  Majoranas at the edges of wires hybridizing into bonding and anti-bonding orbitals when coupled. While, in even number of cases all the Majoranas are coupled pairwise leading to only satellite peaks in the conductance (around $V$= 0), in the case of odd number of wires one Majorana is left unpaired that contributes a conductance of 2 $e^2/h$ at zero bias.

It has been earlier proposed that ZBCP in tunneling experiments can be a smoking gun signature of Majorana modes. However, as it is also known, a ZBCP can originate due to other possible factors as well. That makes ZBCP based detection of Majorana modes somewhat ambiguous. In this context, our proposed experimental scheme is very important. In this scheme, the confirmation of Majorana is not based on the appearance of ZBCP alone, but our calculations provide a detailed scheme based on the number of  transport-active nanowires in a single device where depending on whether the number of wires is odd or even, the ZBCP can be switched $ON$ or switched $OFF$ respectively, in a controlled fashion. Hence, the proposed scheme would provide direct, unambiguous signature of Majorana bound states and topological superconductivity.  

We thank Subhro Bhattacharjee and Tanmoy Das for fruitful discussions. DR thanks DST INSPIRE for financial support. GS acknowledges financial support from Swarnajayanti Fellowship awarded by the Department of Science and Technology, Govt. of India (grant number: \textbf{DST/SJF/PSA-01/2015-16}) to work on the topological phases of matter.



\begin{thebibliography}{100}
		

\bibitem{Hasan}M. Z. Hasan, and C. L. Kane, \textit{Rev. Mod. Phys.}  \textbf{82}, 3045 (2010).


\bibitem{Qi}X. L. Qi, and S. C. Zhang,\textit{ Rev. Mod. Phys.} \textbf{83}, 1057 (2011).
 


\bibitem{Ando} Y. Ando, and M. Sato, \textit{Rep. Prog. Phys.} \textbf{80}, 076501 (2017).


\bibitem{Alicea(a)}J. Alicea, \textit{Rep. Prog. Phys.} \textbf{ 75}, 076501 (2012).


\bibitem{Beenakker}C. W. J. Beenakker, \textit{ Annu. Rev. Con. Mat. Phys.} \textbf{4}, 113 (2013).

\bibitem{Lutchyn}R. M. Lutchyn, J. D. Sau, and  S. Das Sarma, \textit{ Phys. Rev. Lett.} \textbf{ 105},  077001 (2010).

\bibitem{Oreg} Y. Oreg, G. Refael and F. von Oppen, \textit{ Phys. Rev. Lett.} \textbf{ 105}, 177002 (2010).

\bibitem{Mourik} V. Mourik, K. Zuo, S. M. Frolov, S. R. Plissard, E. P. A. M. Bakkers and L. P. Kouwenhoven, \textit{Science} \textbf{336}, 1003 (2012).

\bibitem{Deng} M. T. Deng, C. L. Yu, G. Y. Huang, M. Larsson, P. Caroff, and H. Q. Xu, \textit{Nano Lett.} \textbf{12}, 6414 (2012).

\bibitem{Rokhinson}L. P. Rokhinson, X. Liu, and J. K. Furdyna, \textit{ Nat. Phys.} \textbf{8},  795 (2012).

\bibitem{Anindya} A. Das, Y. Ronen, Y. Most, Y. Oreg, M. Heiblum and H. Shtrikman \textit{ Nat. Phys.} \textbf{ 8},  887 (2012).


\bibitem{Finck} A. D. K. Finck, D. J. Van Harlingen, P. K. Mohseni, K. Jung, and X. Li,   \textit{ Phys. Rev. Lett.} \textbf{ 110},  126406 (2013).


\bibitem{Churchill} H. O. H. Churchill, V. Fatemi, K. Grove-Rasmussen, M. T. Deng, P. Caroff, H. Q. Xu, and C. M. Marcus, \textit{Phys. Rev. B} \textbf{87},  241401 (2013).


\bibitem{Chang} W. Chang, S. M. Albrecht, T. S. Jespersen, F. Kuemmeth, P. Krogstrup, J. Nygrd and C. M. Marcus, \textit{Nat. Nanotechnol.} \textbf{ 10}, 232 (2015).


\bibitem{Kitaev(a)}A. Kitaev, \textit{Phys. Usp.} \textbf{ 44}, 131 (2001).



\bibitem{Read} N. Read, and D. Green, \textit{ Phys. Rev. B} \textbf{ 61}, 10267 (2000).


\bibitem{Alicea(b)}J. Alicea, \textit{Phys. Rev. B}
\textbf{81}, 125318 (2010).

\bibitem{Fu}  L. Fu, and C. L Kane, \textit{Phys. Rev. Lett.} \textbf{ 100}, 096407 (2008).

\bibitem{Nagaosa}D. Asahi, and N. Nagaosa \textit{Phys. Rev. B}\textbf{ 86}, 100504(2012).


\bibitem{Kitaev(b)}A. Kitaev, \textit{Annals of Physics } \textbf{303}, 2 (2003).


\bibitem{Sarma(a)}S. Das Sarma, M. Freedman, and C. Nayak, \textit{Physics Today}\textbf{ 59}, 32 (2006).


\bibitem{Nayak}C. Nayak, S. H. Simon, A. Stern, M. Freedman, and S. Das Sarma, \textit{Rev. Mod. Phys.} \textbf{ 80}, 1083 (2008).


\bibitem{Leijnse}M. Leijnse, and K. Flensberg, 
\textit{Semicond. Sci. Technol.} \textbf{ 27}, 124003 (2012).



\bibitem{Stanescu}T.D. Stanescu, and S. Tewari, \textit{J. Phys. Condens. Matter }\textbf{ 25}, 233201 (2013).


\bibitem{Elliot}S. R. Elliott, and M. Franz, \textit{Rev. Mod. Phys. } \textbf{87}, 137 (2015).


\bibitem{Sarma(b)}S. Das Sarma, M. Freedman, and C. Nayak, 
\textit{Npj Quantum Information} \textbf{1}, 15001 (2015).

\bibitem{Seroussi}I. Seroussi, E. Berg, and Y. Oreg, \textit{Phys. Rev. B} \textbf{89}, 104523 (2014).

\bibitem{Liu} C.-X. Liu, J. D. Sau, T. D. Stanescu, and S. Das Sarma, \textit{Phys. Rev. B} \textbf{96}, 075161 (2017).

\bibitem{Cho} S. Cho, R. Zhong, J. A. Schneeloch, G. Gu, and N. Mason \textit{Sci. Rep.} \textbf{6}, 21767 (2016).

\bibitem{Pikulin} D. I. Pikulin, J. P. Dahlhaus, M. Wimmer, H. Schomerus and C.W.J Beenakker, \textit{New J. Phys. } \textbf{14}, 125011 (2012).

\bibitem{Qu} C. Qu, M. Gong, Y. Xu, S. Tewari, and C. Zhang, \textit{ Phys. Rev. A } \textbf{92}, 023621 (2015).


\bibitem{Blonder}G. E. Blonder, M. Tinkham, and T. M. Klapwijk, \textit{ Phys. Rev. B} \textbf{ 25}, 4515 (1982).


\bibitem{Sarma(c)}S. Das Sarma, J. D. Sau, and T. D. Stanescu, \textit{ Phys. Rev. B} \textbf{ 86}, 220506 (2012).



\bibitem{Blonder}G. E. Blonder, M. Tinkham, and T. M. Klapwijk, \textit{ Phys. Rev. B} \textbf{ 25}, 4515 (1982).


\bibitem{Groth}C. W. Groth, M. Wimmer, A. R. Akhmerov, and X. Waintal, \textit{New J. Phys. } \textbf{16}, 063065 (2014).




\end{thebibliography}
\end{document}